	\documentclass[a4paper,11pt]{article}

	\usepackage[nolist]{acronym}	
	\usepackage[export]{adjustbox}	
	\usepackage{booktabs}	
	\usepackage{comment}
	\usepackage{csvsimple}  
	\usepackage{physics}
	\usepackage{pos}	
	\usepackage{siunitx}
	\usepackage{subcaption}  
	
	\usepackage{hyperref}
	\usepackage{cleveref}	

	\DeclareUnicodeCharacter{2212}{-}	

	\newcommand{\amuhlo}{a_\mu^\mathrm{HLO}}
	\newcommand{\qedc}{\alpha}	
	\newcommand{\qedcu}{\gamma}	
	\newcommand{\dqedc}{\Delta\qedc}
	\newcommand{\dqedchad}{(\dqedc)_\text{had}}
	\newcommand{\pionformfactor}{F_\pi (\omega^2)}
	\newcommand{\svpf}{\hat{\Pi}}
	\newcommand{\weakma}{\sin^2 \theta_W}
	\newcommand{\dweakma}{\Delta \weakma}
	\newcommand{\dweakmahad}{( \dweakma )_\text{had}}

	\begin{acronym}
		\acro{bsm}[BSM]{physics beyond the Standard Model}
		\acro{cls}[CLS]{Coordinated Lattice Simulations}
		\acro{gs}[GS]{Gounaris-Sakurai}
		\acro{hp}[HP]{Hansen-Patella}
		\acro{ib}[IB]{isospin breaking}
		\acro{local}[\textit{l}]{\textit{local}}
		\acro{mll}[MLL]{Meyer-Lellouch-L\"uscher}
		\acro{oet}[OET]{one-end trick}
		\acro{ps}[\textit{ps}]{\textit{point-split}}
		\acro{qed}[QED]{Quantum Electrodynamics}
		\acro{tmr}[TMR]{time-momentum representation}
		\acro{svpf}[sVPF]{subtracted vacuum polarization function}
	\end{acronym}

\title{The~hadronic~contribution to the running of the electromagnetic~coupling and electroweak~mixing~angle}
\ShortTitle{The hadronic contribution to $\dqedc(q^2)$ and $\dweakma(q^2)$}

\author*[c,d,e,f]{Teseo~San~Jos\'e}
\author[a]{Marco~C\`e}
\author[b]{Antoine~G\'erardin}
\author[e,f]{Georg~von~Hippel}
\author[c,d,e,f]{Harvey~B.~Meyer}
\author[c,d,e,g]{Kohtaroh~Miura}
\author[e,f]{Konstantin~Ottnad}
\author[h]{Andreas~Risch}
\author[]{Jonas~Wilhelm}
\author[a,c,d,e,f]{Hartmut~Wittig}

\affiliation[a]{Department of Theoretical Physics, CERN\\
  1211 Geneva 23, Switzerland}
\affiliation[b]{Aix Marseille Universit\'e, Université de Toulon, CNRS, CPT,\\
	Marseille, France}
\affiliation[c]{Helmholtz-Institut Mainz, Johannes Gutenberg-Universität Mainz,\\
	Staudingerweg 18, 55128 Mainz, Germany}
\affiliation[d]{GSI Helmholtzzentrum für Schwerionenforschung,\\
	Planckstraße 1, 64291 Darmstadt, Germany}
\affiliation[e]{PRISMA\textsuperscript{+} Cluster of Excellence, Johannes Gutenberg-Universität Mainz,\\
	Staudingerweg 9, 55128 Mainz, Germany}
\affiliation[f]{Institut für Kernphysik, Johannes Gutenberg-Universität Mainz,\\
	Johann-Joachim-Becher-Weg 45, D 55128 Mainz, Germany}
\affiliation[g]{Kobayashi-Maskawa Institute for the origin of particles and the Universe, Nagoya University Furo-cho,\\
	Chikusa-ku, Nagoya Aichi 464-8602, JAPAN}
\affiliation[h]{John von Neumann-Institut für Computing NIC, Deutsches Elektronen-Synchrotron DESY,\\ Platanenallee 6, 15738 Zeuthen, Germany}

\emailAdd{msanjosp@uni-mainz.de}

\abstract{As present and future experiments, on both the energy and precision frontiers, look to identify new \ac{bsm}, we require more precise determinations of fundamental quantities, like the \acs{qed} and electroweak couplings at various momenta. These can be obtained either entirely from experimental measurements,
or from one such measurement at a particular virtuality combined with
the couplings' virtuality dependence computed within the SM. Thus, a precise, entirely theoretical determination of the running couplings is highly desirable, even more since the preliminary results of the E989 experiment in Fermilab were published.
We give results for the hadronic contribution to the \acs{qed} running coupling $\qedc(Q^2)$ and weak mixing angle $\weakma(Q^2)$ in the space-like energy region $(0, 7]~\text{GeV}^2$ with a total relative uncertainty of $2\%$ at energies $Q^2 \ll \SI{1}{\giga\eV\squared}$, and $1\%$ at $Q^2 > \SI{1}{\giga\eV\squared}$.
\acresetall\vspace*{0.5cm}
\begin{flushright}
  CERN-TH-2021-126\\
  DESY-21-137\\
  MITP/21-038
\end{flushright}
}

\FullConference{%
 The 38th International Symposium on Lattice Field Theory, LATTICE2021
  26th-30th July, 2021
  Zoom/Gather@Massachusetts Institute of Technology
}

\begin{document}
\maketitle

\section{Introduction}

The \acs{qed} running coupling at the Z-pole mass $\qedc(M_Z^2)$ enters in the electroweak global fit. It is also closely related to the leading hadronic contribution to the anomalous magnetic moment of the muon, $\amuhlo$, and both quantities can be obtained using the optical theorem and the R-ratio data, which is the cross-section $\sigma \left( e^+ e^- \rightarrow \text{hadrons} \right)$ normalized by $\sigma \left( e^+ e^- \rightarrow \mu^+ \mu^- \right)$. For an up-to-date R-ratio data collection, see \cite{Davier:2019can,Jegerlehner:2019lxt,Keshavarzi:2019abf}. In the near future, the MUonE experiment \cite{CARLONICALAME2015325,Abbiendi:2016xup} at CERN will compute $\amuhlo$ from the running coupling via a dispersion relation.

The weak mixing angle $\weakma$ is a probe for \acs{bsm} physics, in particular at energies much smaller than the electroweak scale $q^2 \ll M_Z^2$.
It can be measured in neutrino-nucleus scattering, atomic parity violation and parity-violating lepton scattering. However, unlike $\alpha$, it is known with a poor precision in this energy range. The upcoming experiments P2 at MESA \cite{Becker:2018ggl}, as well as MOLLER \cite{MOLLER:2014iki} and SoLID \cite{Chen:2014psa,doi:10.1142/S2010194516600776} at JLab, aim to improve its precision.
Hadronic effects dominate the uncertainty of both quantities, $\qedc (q^2)$ and $\weakma (q^2)$. In this project, we compute these contributions using a similar approach to the one employed to obtain $\amuhlo$ \cite{Gerardin:2019rua}.

\subsection{Main definitions}

The \acs{qed} coupling $\qedc(q^2)$ is parameterized in the \textit{on-shell} scheme as
\begin{equation}
	\label{eq:qed-coupling}
	\qedc(q^2) = \frac{\qedc}{1-\dqedc(q^2)} ,
\end{equation}
where the fine-structure constant is denoted by $\qedc$. The hadronic contribution to $\dqedc(q^2)$ can be expressed in terms of a \ac{svpf},
\begin{equation}
	\label{eq:qed-vpf-connection}
	\dqedchad (q^2) = 4\pi \qedc \svpf^{\qedcu \qedcu} (q^2) .
\end{equation}
In a similar fashion, to leading-order, the $\weakma$ can be given in the on-shell scheme as \cite{Burger:2015lqa,Jegerlehner:1985gq}
\begin{equation}
	\weakma (q^2) = \weakma \left( 1 + \dweakma (q^2) \right) ,
\end{equation}
where $\weakma$ is evaluated at $q^2 \rightarrow 0$ and the hadronic contribution to the running is
\begin{equation}
	\label{eq:wma-vpf-connection}
	\dweakmahad (q^2) = -\dfrac{4\pi\qedc}{\weakma} \svpf^{Z\gamma} (q^2) .
\end{equation}

\subsection{\Acl{tmr}}

Both $\svpf^{\qedcu \qedcu}$ and $\svpf^{Z \qedcu}$ have an integral representation in the space-like region $Q^2 = -q^2$, which is known as the \acf{tmr} \cite{2013PhRvD..88e4502F,2011EPJA...47..148B},
\begin{equation} \label{eq:time-momentum-representation}
	\begin{aligned}
	&\svpf(Q^2) = \int_0^\infty dx_0 G(x_0) K(x_0, Q^2), & &\text{with} & &K(x_0, Q^2) = x_0^2 - \dfrac{4}{Q^2} \sin^2 \left( \dfrac{Qx_0}{2} \right) .
	\end{aligned}
\end{equation}
The kernel may be evaluated at any $Q^2$, but the lattice spacing and its extent limit the range of virtualities that we can reliably handle. The kernel weights the correlator depending on the virtuality input, and virtualities $Q^2 \sim \left( \pi/a \right)^2$ probe the short distances of the \ac{tmr} correlator $G(x_0)$, yielding strong cut-off effects, while virtualities $Q^2 \ll \SI{1}{\giga\eV\squared}$ emphasize the long-distance part of the correlator.
We compute the two-point functions $G^{\qedcu \qedcu}(x_0)$ and $G^{Z \qedcu}(x_0)$, which we decompose using the isospin basis plus the charm-quark component neglecting charm disconnected contributions,
\begin{align}
	\label{eq:electromagnetic-correlator}
	&G^{\qedcu \qedcu} (x_0) =
	G^{33}(x_0)
	+ \dfrac{1}{3} G^{88}(x_0)
	+ \dfrac{4}{9} C^{c, c}(x_0),\\
	\label{eq:Z-gamma-correlator}
	&G^{Z \qedcu} (x_0) =
	\left( \dfrac{1}{2} - \sin^2 \theta_W \right) G^{\qedcu \qedcu}(x_0)
	- \dfrac{1}{6\sqrt{3}} G^{08}(x_0)
	- \dfrac{1}{18} C^{c, c}(x_0).
\end{align}
The isovector $G^{33}(x_0)$, isoscalar $G^{88}(x_0)$ and $G^{08}(x_0)$ parts can be further decomposed into individual flavour components that are directly accessible on the lattice,
\begin{equation}
	\label{eq:isospin-decomposition}
	\begin{gathered}
		G^{33}(x_0) = \dfrac{1}{2} C^{\ell, \ell}(x_0),
		\\
		G^{88}(x_0) =
		\dfrac{1}{6} \left( C^{\ell, \ell}(x_0) + 2C^{s, s}(x_0) + 2D^{\ell - s, \ell - s}(x_0) \right),
		\\
		G^{08}(x_0) =
		\dfrac{1}{2\sqrt{3}} \left( C^{\ell, \ell}(x_0) - C^{s, s}(x_0) + D^{2\ell + s, \ell - s}(x_0) \right) .
	\end{gathered}
\end{equation}
$C^{\ell, \ell}$, $C^{s, s}$ and $C^{c, c}$ are the light, strange and charm quark-connected contributions, respectively. $D^{\ell - s, \ell - s}$ and $D^{2\ell + s, \ell - s}$ are the quark-disconnected pieces, which cancel at the SU(3)-flavour-symmetric point.

\section{Lattice setup}

We employ an extensive set of \acl{cls} ensembles with $N_f=2+1$ flavours, non-perturbatively $\order{a}$-improved Wilson fermions and tree-level L\"uscher-Weisz gauge action \cite{Bruno:2014jqa,Bruno:2016plf}. For the vector currents, we employ two different discretizations in order to further constrain the continuum extrapolation, the \ac{local} and \ac{ps}, whose renormalization and improvement is explained in \cite{Gerardin:2018kpy} ---the latter is a conserved current, and its renormalization is trivial. \Cref{tab:ensembles} lists the ensembles used in this work.
We employ four different lattice spacings and the meson masses range from $M_\pi = M_K \approx \SI{415}{\mega\eV}$ at the SU(3)-flavour-symmetric point to the physical point, along a trajectory where the sum of the bare \textit{u}, \textit{d} and \textit{s} quark masses is constant. The quark loops for the quark-disconnected contribution constitute the most expensive part of the computation. We compute them via a variant of the method proposed in \cite{Giusti:2019kff} combining the one-end trick \cite{McNeile:2006bz} with a combination of the generalized hopping parameter expansion \cite{Gulpers:2013uca} and hierarchical probing \cite{Stathopoulos:2013aci}. The scale is set using $\sqrt{8t_0^\text{phy}} = 0.415(4)(2)~\text{fm}$ \cite{Bruno:2016plf}.
\begin{table}
		\centering
		\csvreader[
			head to column names,
			before reading = \begin{adjustbox}{width=\textwidth},
			after reading = \end{adjustbox},
			tabular = {l r r l l l l l l r r l},
			late after line = \ifcsvstrcmp{\putline}{x}{\\\midrule}{\\},
			table head = \toprule & {$T/a$} & {$L/a$} & {$t_0^\mathrm{sym}/a^2$} & {a [fm]} & {L [fm]} & \multicolumn{2}{l}{$M_\pi$, $M_K$ [MeV]} & $M_\pi L$ & \multicolumn{3}{l}{$\#$ cnfg (con., dis., charm)} \\\midrule,
			table foot = \bottomrule,]
  	{ensemble_specifications.csv}
  	{t=\time, l=\length, a=\lattice}
    {\ifcsvstrcmp{\id}{H105}{\id$^*$}{\ifcsvstrcmp{\id}{H200}{\id$^*$}{\id}} &
     \time &
     \length &
     \ifcsvstrcmp{\putline}{x}{\tsym}{} &
     \ifcsvstrcmp{\putline}{x}{\lattice}{} &
     \lfm &
     \multicolumn{2}{c}{\ifcsvstrcmp{\mpi}{\mK}{\mpi}{\mpi \quad \mK}} &
     \mpl &
     \ncfgconn &
     \ncfgdis &
     \ncfgcharm}
    \caption{Set of \ac{cls} ensembles used in this work. For each one, we give name, geometry, scale in different units, physical size, meson masses and statistics used, differentiating between the connected light and strange contributions, the disconnected pieces and the charm component. We also give $M_\pi L$, as it is related to the size of the finite-size effects present. Ensembles H105 and H200 are only used in the study of the latter, and do not enter the final analysis.}
  	\label{tab:ensembles}
	\end{table}

\subsection{Analysis}

\begin{figure}
		\centering
		\begin{subfigure}{0.42\textwidth}
			\centering
			\includegraphics[width=\textwidth]{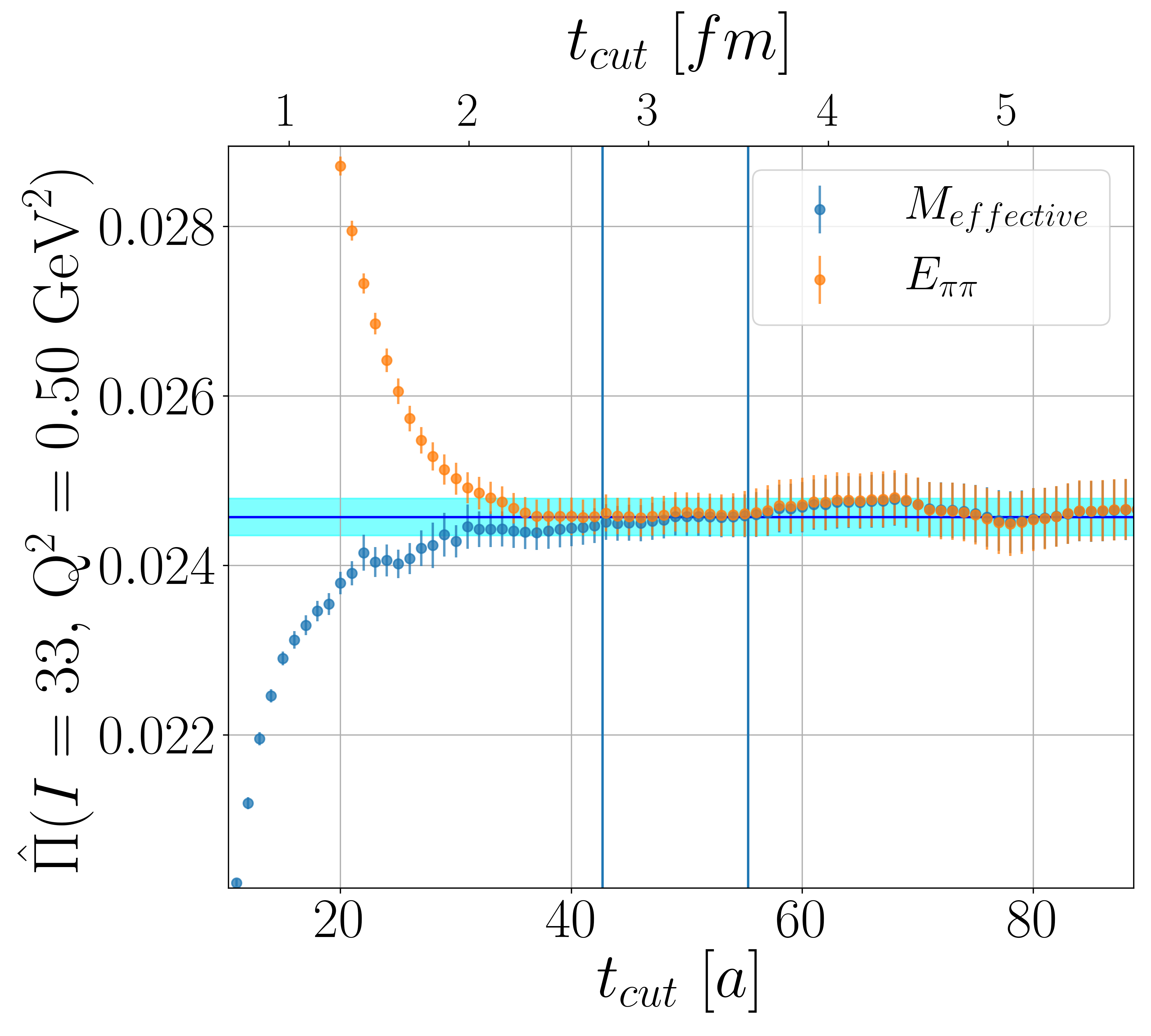}
			\label{fig:E250-33-vv-bounding}
		\end{subfigure}
		\begin{subfigure}{0.42\textwidth}
			\centering
			\includegraphics[width=\textwidth]{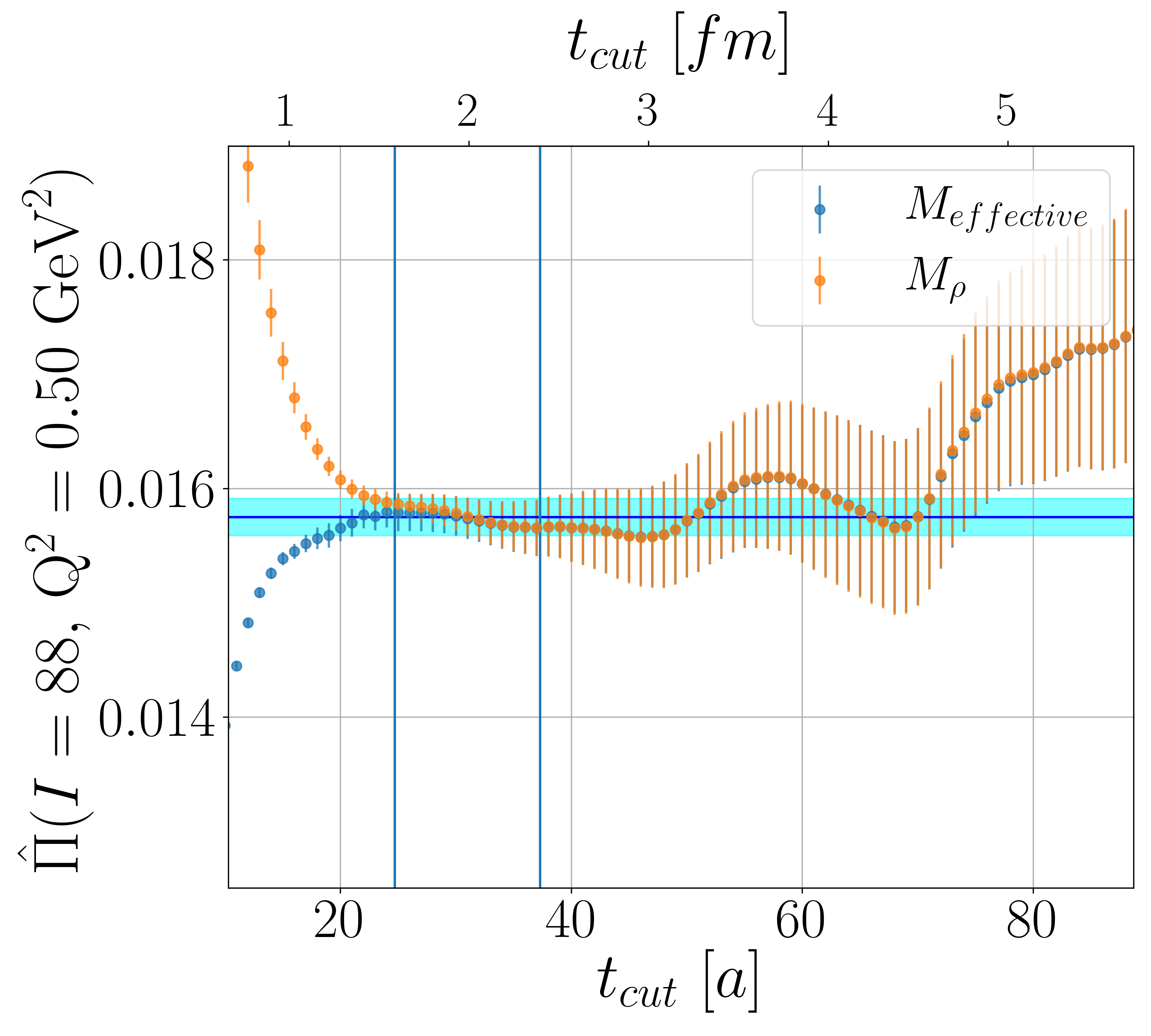}
			\label{fig:E250-88-vv-bounding}
		\end{subfigure}
		\caption{Bounding method applied at the physical pion mass (ensemble E250) and $Q^2 = \SI{0.5}{\giga\eV\squared}$. The data-points indicate the bounds as a function of $t_\text{cut}$, the interval where both are averaged is given between the vertical lines, and the improved estimate is given by the cyan band. LHS: isovector component. RHS: isoscalar contribution.}
		\label{fig:bounding-example}
	\end{figure}

We improve the signal-to-noise ratio by applying the bounding method \cite{Budapest-Marseille-Wuppertal:2017okr} to our correlators. We employ the effective mass $M_\text{eff} (x_0) = \log \left( G(x_0)/G(x_0+1) \right)$ and the ground-state energy $E_0$,
\begin{equation}
	\label{eq:bounding-method}
	\begin{aligned}
	&0 \leq G(x_0^\text{cut}) e^{-M_\text{eff}(x_0) \left( x_0 - x_0^\text{cut} \right)}
	\leq G(x_0)
	\leq G(x_0^\text{cut}) e^{-E_0 \left( x_0 - x_0^\text{cut} \right)} , &
	&x_0 \geq x_0^\text{cut} .
	\end{aligned}
\end{equation}
The ground state is either the $\rho$ meson mass or the two-pion state energy for the isovector contribution. For the isoscalar component, it is either the three-pion state or the $\omega$ meson mass. We estimate the former via its non-interacting energy \cite{Hansen:2020zhy} and the latter employing $M_\rho \approx M_\omega$. We find $M_\rho$ is always lighter and use it as the upper bound. \Cref{fig:bounding-example} shows the result of the bounding procedure at the physical pion mass and energy $Q^2 = \SI{0.5}{\giga\eV\squared}$. On the LHS, we show the isovector component and, on the RHS, the isoscalar. The two sets of points indicate the upper and lower bound, while the two vertical lines show the averaged interval and the cyan band is the improved estimate for $\svpf$.

To estimate the finite-size corrections we employed two different methods, the \ac{hp} procedure \cite{Hansen:2019rbh,Hansen:2020whp} and the \ac{mll} method \cite{Meyer:2011um,Lellouch:2000pv,Luscher:1991cf}, using for the latter the \ac{gs} parameterization of the vector pion form factor $\pionformfactor$ \cite{PhysRevLett.21.244}. \ac{mll} estimates the amplitudes and energies of the finite-size correlator below the inelastic threshold,
\begin{equation}
	G(x_0, L) = \sum_n \abs{A_n}^2 e^{-\omega_n x_0} ,
\end{equation}
where $\abs{A_n}^2$ is related to the module of $\pionformfactor$, and $\omega_n$ to its phase via the L\"uscher equation.
The infinite-volume correlator can be expressed via the continuous spectral representation
\begin{equation}
	\begin{aligned}
		&G(x_0) = \int_0^\infty d\omega \omega^2 \rho(\omega^2) e^{-\omega \abs{x_0}} ,
		&
		&\rho(\omega^2) =
		\dfrac{1}{48\pi^2} \left( 1 - \dfrac{4M_\pi^2}{\omega^2} \right)^{3/2}
		\abs{\pionformfactor}^2
		& 
		x_0 \neq 0 .
	\end{aligned}
\end{equation}
Finally, we take $G(x_0) - G(x_0, L)$ as the finite-size correction for each time-slice $x_0$.
We split the correlator in the time slice $x_i = \left( M_\pi L /4 \right)^2 / M_\pi$, which estimates when the spectral representation of the correlator starts to be described by a handful of states.
We employ the \ac{hp} method for $x_0 < x_i$, and for $x_0 \geq x_i$ both \ac{hp} and \ac{mll} give compatible results.
For two of our ensembles that have the same parameters but the size, H105 and N101, we find a finite-size correction of $\sim 2\%$ for the first, and of $\sim 0.2\%$ for the second, at $Q^2 = \SI{1}{\giga\eV\squared}$.

\begin{figure}
		\centering
		\begin{subfigure}{0.42\textwidth}
			\centering
			\includegraphics[width=\textwidth]{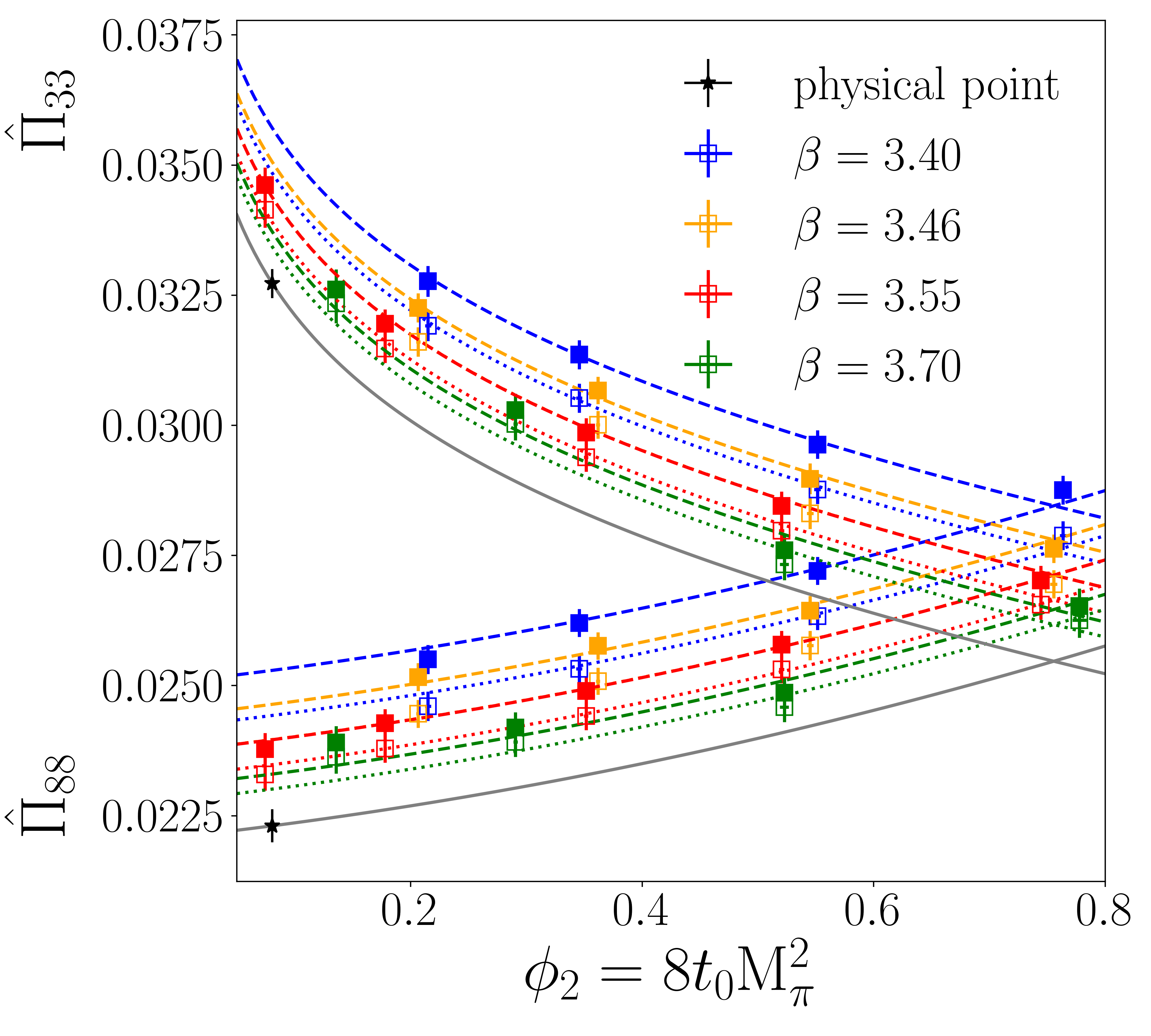}
			\label{fig:33-88-extrapolation}
		\end{subfigure}
		\begin{subfigure}{0.42\textwidth}
			\centering
			\includegraphics[width=\textwidth]{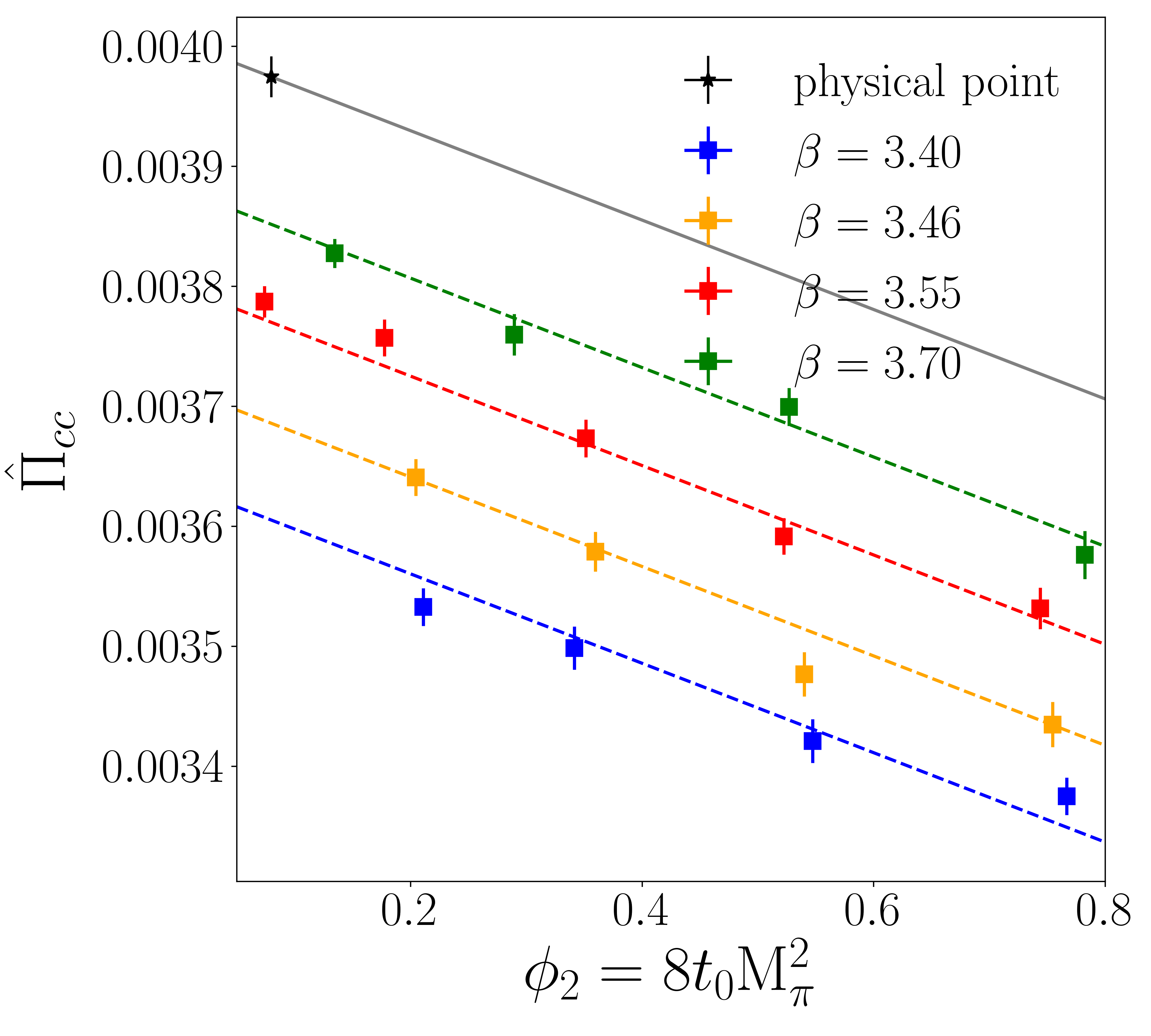}
			\label{fig:c-extrapolation}
		\end{subfigure}	
		\caption{Extrapolation to the physical point at $Q^2 = \SI{1}{\giga\eV\squared}$. The isovector and isoscalar components are represented on the LHS and the charm contribution on the RHS.}
		\label{fig:extrapolation-example}
	\end{figure}

We perform an extrapolation to the physical point $a \rightarrow 0 $, $M_\pi^\text{phy} = \SI{134.9768(5)}{\mega\eV}$, $M_K^\text{phy} = \SI{495.011(15)}{\mega\eV}$, where we employ the meson masses without \ac{ib} effects \cite{criteriawp2021,Zyla:2020zbs}, since they are absent in our calculation. We employ the dimensionless variables \cite{Bruno:2016plf} $a^2/8t_0^\text{sym}$, $\phi_2 = 8t_0 M_\pi^2$, and $\phi_4 = 8t_0 \left( M_\pi^2/2 + M_K^2 \right)$. We extrapolate the isovector and isoscalar components together, as they coincide in the SU(3)-flavour-symmetric point, where in fact $\phi_4^\text{sym} = 3\phi_2^\text{sym}/2$. The fit model of each isospin and correlator discretization can be divided into lattice- and mass-dependent terms, $\svpf(a, \phi_2, \phi_4; d, i) = \svpf_\text{continuum}(a; d) + \svpf_\text{mass}(\phi_2, \phi_4; i)$,
\begin{align}
	&\svpf_\mathrm{continuum}(a; d) =
	\begin{cases}
		\svpf^\mathrm{sym} + \alpha_{2, d} \left( a^2/8t_0^\text{sym} \right) , &
		Q^2 \lesssim \SI{2.5}{\giga\eV\squared}, \\
		\svpf^\mathrm{sym}
		+ \alpha_{2, d} \left( a^2/8t_0^\text{sym} \right)
		+ \alpha_{3, d} \left(a^2/8t_0^\text{sym}\right)^{3/2} , &
		Q^2 \gtrsim \SI{2.5}{\giga\eV\squared},
	\end{cases}
	\\
	&\svpf_\mathrm{mass} (\phi_2, \phi_4; i) =
 	\beta_{1, i} \left( \phi_2 - \phi_2^\mathrm{sym} \right) +
 	\delta \left( \phi_4 - 3\phi_2^\mathrm{sym}/2 \right) +
	\begin{cases}
 		\beta_{2, 33} \log \left( \phi_2/\phi_2^\mathrm{sym} \right) , & i=33 , \\
 		\beta_{2, 88} \left( \phi_2 - \phi_2^\mathrm{sym} \right)^2 , & i=88 .
 	\end{cases}
\end{align}
The subscript $d=\ac{ps}, \ac{local}$ runs over the correlator discretizations and $i=33, 88$ over both isospins. Therefore, there is one pair of $\alpha_2$ and $\alpha_3$ parameters per correlator discretization, and one set of parameters $\beta_1$ and $\beta_2$ per isospin. Also, the curvature of the data is described differently for the isovector and isoscalar components.  \Cref{fig:extrapolation-example} shows the extrapolation to the physical point for the isovector, isoscalar and charm components at $Q^2 = \SI{1}{\giga\eV\squared}$. The different colours indicate the various lattice spacings, the dashed and dotted lines refer to the two correlator discretizations. The grey line is the continuum extrapolation, with the black dot indicating the result at the physical point.
For the charm contribution, the \acl{local} discretization shows much larger artefacts, therefore we only extrapolate the \ac{ps} data. The model employed is $\svpf (a, \phi_2; \text{charm}) = \svpf^\text{phy} +	\alpha_{2, \ac{ps}} \left( a^2/8t_0^\text{sym} \right) + \beta_{1,\text{charm}} \left(\phi_2 -\phi_2^\mathrm{phy} \right)$.
The 08 component is found to be rather insensitive to the lattice spacing, and thus we only study the \ac{ps} discretization. Besides, \cref{eq:isospin-decomposition} anticipates this contribution to be zero at the SU(3)-flavour-symmetric point. It suffices to use the one-parameter model $\svpf(\phi_2, \phi_4; 08) = \svpf_{08} \left( \phi_4 - 3\phi_2/2 \right) =  \svpf_{08} 8t_0 \left( M_K^2 - M_\pi^2 \right)$. To assess the extrapolation's systematic error, we exclude all ensembles with pion masses $M_\pi > \SI{400}{\mega\eV}$, corresponding to approximately $\phi_2 > \num{0.6}$ in \cref{fig:extrapolation-example}. In the case of the charm and 08 components, we also cut all masses $M_\pi > \SI{300}{\mega\eV}$, or $\phi_2 > \num{0.4}$. The fit quality at $Q^2 = \SI{1}{\giga\eV\squared}$ is $\chi^2/\text{ddof} = 1.35$ for the isovector and isoscalar, $\chi^2/\text{ddof}=0.66$ for the 08 term, and $\chi^2/\text{ddof} = 2.5$ for the charm contribution. For the latter, the mass cuts significantly improve the fit quality to $\chi^2/\text{ddof} = 0.85$ without changing the result at the physical point, suggesting the bad fit quality is an artifact of the covariance matrix.

\section{Results}

\begin{figure}
	\centering
	\begin{subfigure}{0.42\textwidth}
		\centering
		\includegraphics[width=\textwidth]{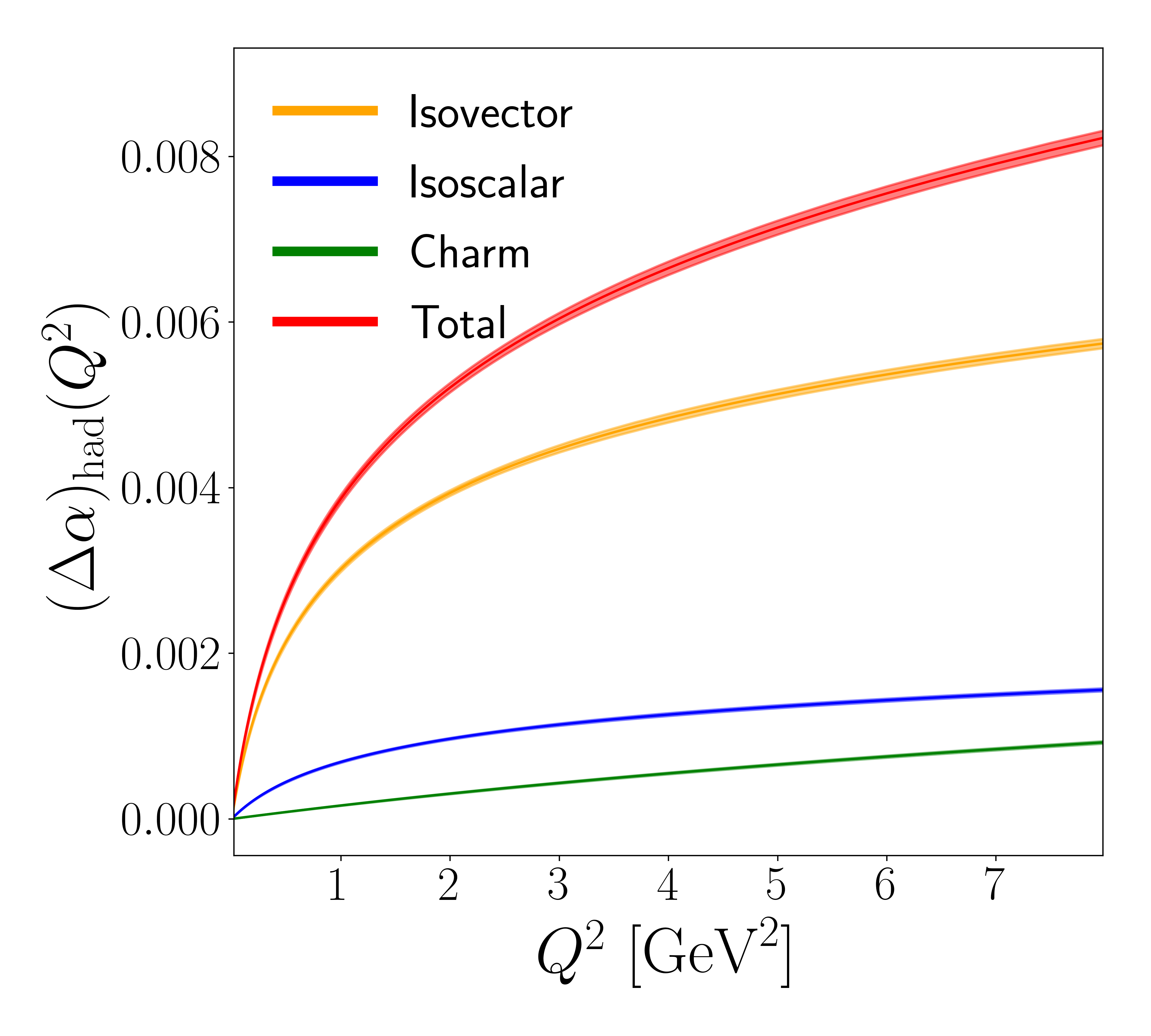}
	\end{subfigure}
	\begin{subfigure}{0.42\textwidth}
		\centering
		\includegraphics[width=\textwidth]{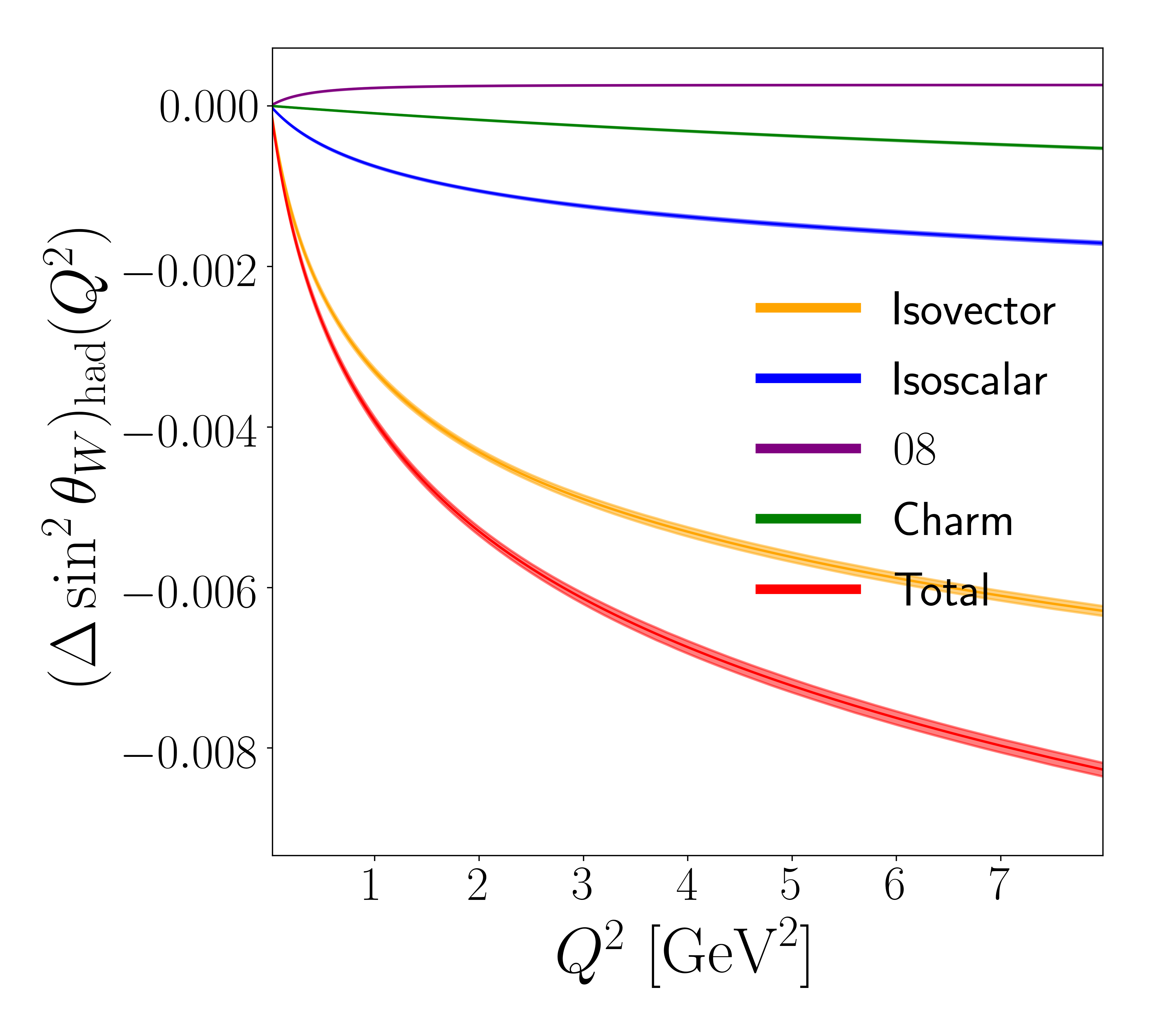}
	\end{subfigure}	
	\caption{Results for \cref{eq:qed-vpf-connection} (LHS) and \cref{eq:wma-vpf-connection} (RHS) at the physical point. We differentiate between the various contributions according to \cref{eq:electromagnetic-correlator,eq:Z-gamma-correlator}, including all prefactors.}
	\label{fig:running}
\end{figure}

\Cref{fig:running} shows our preliminary results for the hadronic contributions at the physical point of both, the \acs{qed} running coupling and the weak mixing angle, defined in \cref{eq:qed-vpf-connection,eq:wma-vpf-connection} respectively. Following \cref{eq:isospin-decomposition}, we present the result of each isospin component, including e.g.~the prefactors 1/3 and 4/9 of \cref{eq:electromagnetic-correlator} to emphasize the
light-quark correlator's dominance. The width shows our total error budget.
The \acs{qed} and strong isospin breaking corrections have been computed for a small subset of ensembles \cite{Risch:2017xxe,Risch:2018ozp,Risch:2019xio}. We use them to estimate the systematic uncertainty due to the absence of \ac{ib} effects in our main computation. The error size for $\dqedchad$ varies between $1\%$ at small energies and $0.5\%$ at the highest momentum probed while, in the case of $\dweakmahad$, it is one order of magnitude smaller, between $0.1\%$ and $0.05\%$, respectively.
The scale setting error, which enters through $8t_0^\text{phy}Q^2$ in the kernel $K(x_0, Q^2)$ of \cref{eq:time-momentum-representation} and the definition of the physical point, is estimated to be $\sim 0.5\%$, using bootstrap sampling.

\section{Conclusions}

In this work, we compute the hadronic contribution to the \acs{qed} running coupling $\qedc(Q^2)$ and weak mixing angle $\weakma(Q^2)$ in the space-like energy region $(0, 7]~\text{GeV}^2$. Our main error sources stem from statistics, the scale setting, the extrapolation to the physical point and the isospin breaking corrections, giving a total relative uncertainty of $2\%$ at energies $Q^2 \ll \SI{1}{\giga\eV\squared}$, and $1\%$ for the range $Q^2 > \SI{1}{\giga\eV\squared}$. A preliminary comparison shows rough agreement between our determination of $\alpha (Q^2)$ and the Lattice results \cite{Burger:2015lqa,Budapest-Marseille-Wuppertal:2017okr}. Currently, we aim to compare our \acs{qed} running coupling with the phenomenological determinations \cite{Davier:2019can,Jegerlehner:2019lxt,Keshavarzi:2019abf}, as well as our $\weakma$ with \cite{Jegerlehner:2019lxt,Jegerlehner:2011mw}. The \acs{qed} and strong isospin breaking effects have been computed on a subset of ensembles so that we may estimate the uncertainty due to their absence in the main computation, but the analysis will be extended to other ensembles in the future \cite{Risch:2019xio,Risch:2018ozp,Risch:2021pos}. We plan to express \cref{fig:running} in an analytic form, employing a rational expression for the running. \cite{KohtarohLattice2021} shows how our results affect $\dqedc (M_Z^2)$.

\bibliography{bibliography}
\bibliographystyle{JHEP}

\end{document}